\documentstyle[twocolumn,pra,aps,epsf]{revtex}      

\begin{document}

\draft
\title{Dielectronic Recombination of Ground-State and
Metastable Li$^{+}$ Ions} 

\author{A. A.~Saghiri,$^{1}$
J.~Linkemann,$^{1}$
M.~Schmitt,$^{1}$
D.~Schwalm,$^{1}$ 
A.~Wolf,$^{1}$
T.~Bartsch,$^{2}$ 
A.~Hoffknecht,$^{2}$ 
A.~M\"uller,$^{2}$ 
W.~G.~Graham,$^{3}$ 
A.~D.~Price,$^{4}$ 
N.~R.~Badnell,$^{4}$ 
T.~W. Gorczyca,$^{5}$ 
J.~A.~Tanis$^{5}$}

\address{$^1$Max-Planck-Institut f\"ur Kernphysik, Heidelberg, and 
Physikalisches Institut der Universit\"at, 
D-69029 Heidelberg, Germany} 

\address{$^2$Institut f\"ur Kernphysik, Strahlenzentrum der 
Justus-Liebig-Universit\"at, D-35392 Giessen, Germany} 

\address{$^3$Department of Pure and Applied Physics, Queen's
  University, Belfast BT7 1NN, Northern Ireland}

\address{$^4$Department of Physics and Applied Physics, University of 
Strathclyde, Glasgow G4 0NG, United Kingdom} 

\address{$^5$Department of Physics, Western Michigan University, 
Kalamazoo, Michigan 49008, USA} 

\date{July 19, 1999 --- {\sl Submitted to Phys. Rev. A}}
\maketitle

\begin{abstract}
  Dielectronic recombination has been investigated for $\Delta n=1$
  resonances of ground-state Li$^{+}(1s^2)$ and for $\Delta n=0$
  resonances of metastable Li$^{+}(1s2s\,^3S)$.  The ground-state
  spectrum shows three prominent transitions between 53 and 64~eV,
  while the metastable spectrum exhibits many transitions with
  energies $<3.2$~eV.  Reasonably good agreement of R-matrix, $LS$
  coupling calculations with the measured recombination rate
  coefficient is obtained.  The time dependence of the recombination
  rate yields a radiative lifetime of $52.2\pm5.0$ s for the 2\,$^3S$
  level of Li$^+$.
\end{abstract}


\narrowtext

In collisions between ions and electrons, recombination \cite{RAI} can
occur either by radiative recombination (RR) (inverse of the
photoelectric effect), or by dielectronic recombination (DR).  This
process starts by the resonant capture of a free electron, mediated by
its interaction with a bound electron (inverse of an Auger
transition), and then proceeds by a radiative transition to a stable
recombined system.  Dielectronic recombination is of fundamental
interest because it gives insight into electron correlation, and of
applied interest for the understanding of astrophysical and laboratory
plasmas.  Extensive experimental studies providing significant tests
of the theory \cite{RAI} have been conducted for H-like, He-like, and
Li-like ions heavier than about carbon.  For such systems,
perturbative techniques can be successfully used in theoretical
calculations because the electron--electron interaction is relatively
weak compared to the electron--nucleus interaction.  In contrast, for
lighter ions perturbative techniques cease to produce results at the
same level of accuracy.  Hence, high quality data for light systems
are required to test the reliability of theoretical predictions.

For the lightest H-like ion, He$^+$, DR was studied previously by
three groups \cite{john,tan,dewi} with successively higher energy
resolution, and through a consequent feedback between experiment and
theory good mutual agreement could be obtained.  For the lightest
He-like ion, Li$^{+}$, the two-electron configuration places
additional constraints on the theory of RR and DR. Moreover, the
existence of relatively long-lived metastable $1s2s\,^1S$ and
$1s2s\,^3S$ states with natural lifetimes of about 0.5 ms and 50 s,
respectively, requires recombination for both the ground state and the
metastable state (especially the long-lived triplet state) to be
understood.  To date, only low-resolution DR measurements for Li$^{+}$
have been carried out \cite{zvod}; the integrated ground-state and
metastable recombination rates could be extracted, but more detailed
comparisons between experiment and theory for specific excited-state
configurations were not possible.

Here we report a high-resolution investigation of DR in Li$^{+}$.
Individual intermediate-excited state configurations contributing to
DR could be resolved for ground-state Li$^{+}(1s^2\,^1S)$ ions, where
DR occurs via $\Delta n=1$ resonances ($1s^2+e \rightarrow 1s2ln'l'$)
at relative electron--ion energies of 53--64 eV, as well as for
metastable Li$^{+}(1s2s\,^3S)$ ions which give rise to $\Delta n=0$
resonances ($1s2s+e \rightarrow 1s2ln'l'$) below 3.2 eV.  The
metastable spectrum is found to be particularly rich with many
closely-spaced transitions.  The time dependence of DR from metastable
ions was also used to determine the radiative lifetime of the 2\,$^3S$
level.  Moreover, RR has been observed for relative electron--ion
energies close to zero.

At the Heidelberg ion-storage ring and electron cooler facility (TSR)
a beam of 13.3 MeV $^7$Li$^{+}$ ions was produced in a tandem
accelerator and injected into the ring.  Using multiturn stacking
injection, which lasted about 0.1~ms, a typical circulating ion
current of 5~$\mu$A was obtained.  While stored, the ions passed
through the collinear electron beam of the electron cooler.  First, by
matching the electron and ion velocities at a laboratory electron
energy of 1.04 keV (`cooling energy') the ion beam was phase-space
cooled to low ($\sim$0.3~mrad) divergence and $\sim$10$^{-4}$ relative
longitudinal momentum spread within about 10~s.  Recombination of the
Li$^{+}$ ions was then measured as a function of the energy $E$
corresponding to the average relative velocity between the electrons
and the ions (referred to as relative energy) by detuning the
laboratory electron energy from the cooling energy by a variable
amount.  In the measurements, the relative energy was alternately set
(on a time scale of milliseconds) first to $E\approx140$ eV to
determine the background recombination rate, then to a desired value
in the energy ranges of ground- or metastable-state recombination, and
then back to the cooling energy.

Those ions which recombined to form neutral Li are no longer deflected
by the storage ring dipole magnet following the electron cooler, and
can thus be counted with a surface-barrier detector behind this
magnet.  Field ionization in the motional electric field experienced
by the fast Li atoms in the dipole magnet introduces a principal
quantum number cutoff \cite{kanter} of $n_{c}=7$ so that atoms
reaching the magnet in states with $n>n_{c}$ are not detected.

The energy resolution was determined by the electron temperatures of
$kT_\perp=18$ meV and $kT_\parallel=0.2$ meV \cite{pas}, which
resulted in a FWHM energy spread of $\sim$0.02 eV in the electron--ion
center-of-mass (c.m.)\ frame at $E\lesssim0.2$ eV, increasing to
$\sim$0.4 eV at $E=60$~eV.  The electron current was 35 mA in the
lower and 60 mA in the upper energy range, respectively, corresponding
to electron densities $n_e$ of 5 and $8\times10^6$ cm$^{-3}$.

Recombination was measured in a time interval of 12--70 s after each
injection, the total cycle time being close to the $1/e$ storage
lifetime of $\sim$60 s determined by the ground-state stripping rate
in the rest gas, $\lambda_{g}^s$.  Measurements were repeated for
successive injections over several hours.  From earlier experiments
\cite{schroe} employing laser cooling of $^7$Li$^+$ ions on the
$2\,^3S$-$2\,^3P$ transition, a substantial initial fraction
$f_m^0\approx0.1$--0.3 of metastable $2\,^3S$ ions is known to exist
in the TSR under these running conditions.  The unknown initial
fraction of metastable singlet ($2\,^1S$) ions is expected to have
decayed by the time the data were recorded.  The decay constant of the
triplet ($2\,^3S$) metastable Li$^+$ ions stored in the ring was
$\sim$(19\,s$)^{-1}$, given mainly by the sum of the rest-gas
stripping rate $\lambda_{m}^{s}$ and the radiative decay rate
$\lambda_{mg}^{\gamma}$ of the metastable ions.  Stripping of $2\,^3S$
ions is expected \cite{nikolaev} to occur faster than that of $1\,^1S$
ions because of the lower ionization potential of the triplet state.
The energy region of metastable DR ($E=0$--3.5 eV) was scanned in the
two time intervals of 12--42 and 54--70 s after injection, while the
ground-state DR spectrum ($E=52$--64 eV) was measured between 42--54
s.  Details concerning the determination of the relative energy and
the c.m.\ electron energy distribution can be found in Ref.\ 
\cite{kilgus}.  The accuracy of the relative-energy scale is estimated
to be $\pm 0.02$ eV at 1 eV and $\pm 0.09$ at 60 eV.

With the recombination rate coefficients $\alpha_g(E)$ and
$\alpha_m(E)$ for ground and metastable ions, respectively, and the
metastable ion fraction $f_m(t)$, the measured, background subtracted
recombination rate is given by
\begin{eqnarray}
R(E,t) &=& \eta \gamma^{-2}  \cdot N_i(t) \cdot n_e\nonumber \\ 
& &{}\times \left[ \Bigl(1-f_m(t)\Bigr) \alpha_g(E) 
+ f_m(t)\alpha_m(E) \right]\;,
\label{eq:rates}
\end{eqnarray}
where $\gamma=1.002$ is the Lorentz factor due to the relativistic
transformation between the ion rest frame and the laboratory frame,
and $\eta$ the ratio between the interaction length (1.5~m) and the
ring circumference (55.4~m).  The number of stored ions $N_i(t)$ was
determined from the circulating beam current measured by a
non-destructive current probe as well as by counting events at a beam
profile monitor \cite{hochadel}.

\begin{figure}[f]
\epsfxsize=6.8cm
\centerline{\epsffile{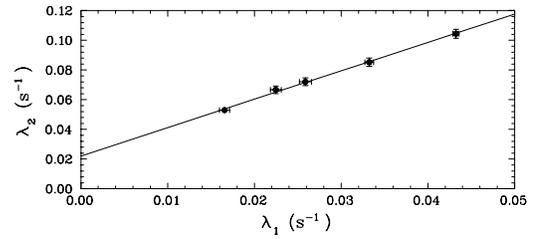}}
\vspace{1mm}
\caption[]{Decay constants representing the total decay rates
  $\lambda_1$ of ground-state and $\lambda_2$ of metastable Li$^+$
  ions, as measured for five residual gas pressures.  Extrapolation to
  $\lambda_1=0$ (corresponding to vanishing pressure) yields
  $\lambda_{mg}^{\gamma} + \lambda_{m}^e=0.0221\pm0.0015$ s$^{-1}$.}
\label{f_extrap}
\end{figure}

In addition to the energy scans, the time dependence of the
recombination rate on selected DR peaks was measured, performing a
sequence of fast scans over resonances at $E=0.14$ and 1.2 eV
(representing mainly metastable-ion DR) and 54 eV (representing only
ground-state DR); simultaneously the beam current signals from the
current probe and the beam profile monitor as well as the count rates
at $E=0$ and at $E=140$ eV were recorded.  The scan sequence was
repeated 25 times after each injection, covering storage times of
12--75 s.  The integrated count rates were fitted jointly as functions
of time by two-component exponential decay curves, varying the two
decay constants common to all curves and appropriate separate
amplitudes.  Under the assumption (justified below) that collisional
re-feeding of the metastable state by rest gas collisions can be
neglected, the two resulting time constants are $\lambda_1 =
\lambda_{g}^s$ and $\lambda_2 = \lambda_{m}^{s} + \lambda_{mg}^{q} +
\lambda_{mg}^{\gamma} + \lambda_{m}^e$, where $\lambda_{mg}^{q}$
denotes the collisional quenching rate and $\lambda_{m}^e$ is the
electron impact ionization rate \cite{berrington} of metastable ions
in the electron cooler occuring while the relative energy is above the
2\,$^3S$ ionization threshold (the same effect for ground-state ions
is negligible).  Measurements of the time variation at five residual
gas pressures differing by factors of up to $\sim$3 yielded for
$\lambda_{1,2}$ the results shown in Fig.\ \ref{f_extrap}.  As
$\lambda_{g}^s$, $\lambda_{m}^s$, and $\lambda_{mg}^{q}$ all vary
proportional to the rest gas pressure, $(\lambda_{m}^{s} +
\lambda_{mg}^{q}) / \lambda_{g}^{s}$ can be deduced from the slope of
the straight line through these data, while the rest-gas independent
decay rate $\lambda_{mg}^{\gamma} + \lambda_{m}^e$ of the metastable
state follows from linear extra\-polation to $\lambda_{1}=0$.  After
subtraction of the estimated \cite{berrington} electron impact
ionization rate of $\lambda_{m}^e=0.003\pm0.001$ s$^{-1}$ the
radiative decay rate of $\lambda_{mg}^{\gamma}=0.0191\pm0.0018$
s$^{-1}$ is obtained, which corresponds to a natural lifetime of
$\tau_{M1}=52.2\pm5.0$ s (1$\sigma$ error) for the 2\,$^3S$ level.
This result matches well the existing theoretical predictions
\cite{drake,devrianko} of 49.04--49.14 s as well as the only previous
experimental result of $\tau_{M1}=58.6 \pm 12.9$ s (2$\sigma$ error)
\cite{knight}.  The fact that the slope of $\lambda_2$ vs.\ 
$\lambda_1$ lies close to the expected ratio of the metastable and
ground-state rest-gas stripping rates ($\sim$2 \cite{nikolaev})
indicates that the collisional quenching rate was small.

After normalization to the ion number $N_i(t)$ the time-dependent rate
in the 54-eV peak is proportional to $1-f_m(t)$ [see Eq.\ 
(\ref{eq:rates})], while the rates in the 0.14- and 1.2-eV peaks
dominantly vary as $f_m(t)$ with a small additional contribution from
ground-state RR that could be estimated.  Fits to the normalized rates
in the 54-eV peak yielded only rather inaccurate results for the
initial metastable fraction $f_m(0)=f_m^0$, scattering between 0.1 and
0.4, because the change of the ground-state fraction $1-f_m(t)$ during
the measuring time was only slightly larger than the statistical
counting errors.  Corresponding fits to the low-$E$ peaks (focusing on
the signal decay at longer storage times) allowed the rate of
collisional re-feeding of the 2\,$^3S$ state to be estimated to be
$<$0.02 of the ground-state stripping rate $\lambda_g^s$, justifying
the assumption made above within the statistical error limits.

The rates $R(E,t)$ were converted to the experimental rate
coefficients $\alpha_g(E)$ and $\alpha_m(E)$ using Eq.\ 
(\ref{eq:rates}); appropriate rate equations together with the
measured time constants $\lambda_{1,2}$ yielded the functional
dependence $f_m(t)$.  In the low energy range (0--3.5 eV) the measured
rates contain contributions to DR and to RR from metastable ions
[$\alpha_m(E)$] and to RR from ground-state ions [$\alpha_g(E)$].
They were extracted separately combining the data from the two storage
time intervals in which this energy region was scanned.  The
experimental recombination rate coefficient $\alpha_m(E)$ for
metastable Li$^+$ ions is shown in Fig.\ \ref{ms}.  Also shown is the
theoretical rate coefficient (see below) for both DR and RR convoluted
with the experimental electron velocity distribution.  The intrinsic
determination of the metastable fraction being too inaccurate, it was
set to $f_m^0=0.19$, which yields the best agreement of the data
between 0.25 and 0.35 eV (see bottom part of figure) with the
theoretical prediction.  Thus, $f_m^0$ is adjusted to fit the
calculated {\em non-resonant} (RR) rate of metastable Li$^+$.
Different values of $f_m^0$ would essentially scale the experimental
result for $\alpha_m(E)$ as $\propto 1/f_m^0$.  The value of
$f_m^0=0.19$ is compatible with the observed time dependences, as well
as with the values estimated from earlier laser cooling experiments
\cite{schroe}.  The systematic error in the normalization to the ion
and electron beam intensity is estimated to $\pm20\%$; however, in
view of the uncertainty in $f_m^0$ we attribute an overall systematic
error of $\pm 50\%$ to $\alpha_m(E)$.  The resulting smooth
ground-state RR rate coefficient $\alpha_g$ reaches a value of
$(2\pm1)\times10^{-12}$ cm$^3$\,s$^{-1}$ at near-zero relative energy
and drops to $(2\pm1)\times10^{-13}$ cm$^3$\,s$^{-1}$ at $E=0.3$ eV.

\begin{figure}[f]
\epsfxsize=6.8cm
\centerline{\epsffile{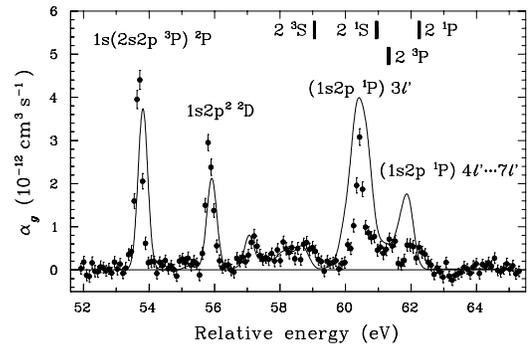}}
\vspace{1mm}
\caption[]{Dielectronic recombination rate coefficient $\alpha_g(E)$
  for ground-state Li$^{+}\,(1s^2)$ ions.  Dots: experimental data;
  line: theory for $n_{c}=7$.  Excitation thresholds from the
  $1\,^1S$ state are indicated.}
\label{gr}
\end{figure}

In the high energy range (52--65 eV) only $\Delta n=1$ DR of
ground-state ions occurs, the contribution from the RR of metastable
and ground-state ions being negligible.  ($\Delta n=1$ DR of
metastable Li$^+$ ions, leading to triply excited Li, lies in
different energy regions not scanned in this experiment.)  The DR rate
coefficient $\alpha_g(E)$ for ground-state Li$^+$ ions is shown in
Fig.~\ref{gr}.  The uncertainty in $f_m^0$ here yields only a small
contribution to the systematic error since the metastable fraction at
the time of measurement is already down to $\lesssim0.04$.  Hence, a
systematic error of $\pm 20\%$ is estimated for $\alpha_g(E)$.

To obtain the theoretical rate coefficients shown in Figs.~\ref{ms}
and \ref{gr} unified photorecombination cross sections representing
both DR and RR were computed in $LS$ coupling using a radiation damped
R-matrix approach \cite{rob,gorc}.  In fact, this case is one of the
few where damping is negligible \cite{badn}, and we obtained the same
results by applying detailed balance to undamped R-matrix
photoionization cross sections.  To compare the calculations with
experiment, contributions to DR where one of the electrons occupies a
configuration with $n'>n_c=7$ were excluded from the calculations.
The $\Delta n=0$ DR spectrum of metastable Li$^+$ ions (Fig.~\ref{ms})
is dominated by $n'$-manifolds converging to the $2\,^3S$-$2\,^1P$
excitation threshold at 3.2 eV [configurations $1s2p(^1P)n'l'$].  At
low energies terms of the configuration $1s2p(^3P)3l'$ also appear.
There is good agreement between theory and experiment up to 0.6 eV,
although the theoretical line energies seem to be too high by
$\sim$0.03 eV and the double structure of the peak at 0.14 eV is not
reproduced.  At higher energies the theory appears to overestimate the
DR cross section significantly, but still yields a resonable
representation of the general spectral shape.

It appears that with $n_{c}=7$ we do not account for the recombination
rate observed above 3 eV.  This is probably due to Li atoms formed
with $n>n_{c}$, cascading to lower $n$ states during the
time-of-flight (0.3 $\mu$s) prior to the dipole magnets.  Furthermore,
the sharp cutoff $n_{c}$ represents only an approximate description of
the field ionization process \cite{kanter}.  Particularly large
discrepancies between theory and experiment occur above the
$2\,^3S$-$2\,^3P$ excitation threshold (2.28 eV).  Here, states with
high angular momentum $l'\geq3$ strongly contribute to the calculated
DR rate, as demonstrated by eliminating these states from the
calculation (lower, dashed curve in Fig.~\ref{ms}).  Using this $l$
cutoff brings the calculated DR rate much closer in magnitude to the
experiment.  The possibility that Li atoms in high $l$ states (formed
by radiative stabilization of high-$l$ resonances) suffer field
ionization even below the cutoff $n_{c}$ can probably be ruled out.  A
possible experimental explanation for these discrepancies could arise
from stray electric fields in the interaction region, estimated to be
$\lesssim3$ V/cm; the DR rate above $\sim$1.9 eV is reduced by
autoionization into additional open continua ($1s2s\,^1S$,
$1s2p\,^{3}P$) and, as remains to be verified in detail, high-$l$
resonances may assume larger autoionization rates than calculated
through $l$-mixing in the stray electric fields.

Regarding the ground-state DR (Fig.~\ref{gr}) experiment and theory
are in good agreement for the $1s2l2l'$ configurations near 54 and 56
eV.  These terms are well known from ion--atom collision
\cite{ziem,rdb} and photoionization studies \cite{kierk,kierl}.  Along
the $1s2p(^1P)n'l'$ series one higher resonance, probably from
$1s2p(^1P)3l'$, is observed near 60 eV, but higher series members
apparently are reduced by autoionization into the $1s2l$ continua.
Near 60~eV the energy-integrated theoretical rate coefficient,
representing several unresolved DR resonances, exceeds the
experimental value (apparently a single resonance) by about a factor
of 3.  Also, the calculations predict a higher peak near 62 eV due to
Rydberg states with $n=4$--7.

In summary, Li$^+$ ions with a strong population of the long-lived
2\,$^3S$ state were stored for times on the order of the natural
lifetime of the metastable level.  This lifetime was determined with
an error of $\pm10\%$, confirming the theoretical predictions.
Experimental DR rate coefficients were obtained separately for the
metastable and the ground state.  DR from the 2\,$^3S$ ions shows a
rich spectrum of lines from doubly excited, higher angular momenta
configurations of Li that cannot be reached by photoionization from
the Li ground state and have not been observed before; their
description presents a significant and important challenge to the
theory of this fundamental two-electron system.  An R-matrix
calculation reproduces the spectral shape reasonably well.  The
ground-state DR spectrum shows three prominent and several weaker
lines at energies of 54--62 eV, and there is reasonable agreement with
theoretical calculations except for lines near the $1s2l$ excitation
thresholds ($>60$~eV).

This work was supported by the Human Capital and Mobility Programme of
the European Community, and by the German Ministry of Education,
Science, Research, and Technology (BMBF) under contracts no.\ 
06\,HD\,854I and 06\,GI\,848.  J.A.T. was supported in part by the
Division of Chemical Sciences, Office of Basic Energy Sciences, Office
of Energy Research, U.S.  Department of Energy.

\begin{figure}[f]
\epsfxsize=6.8cm
\centerline{\epsffile{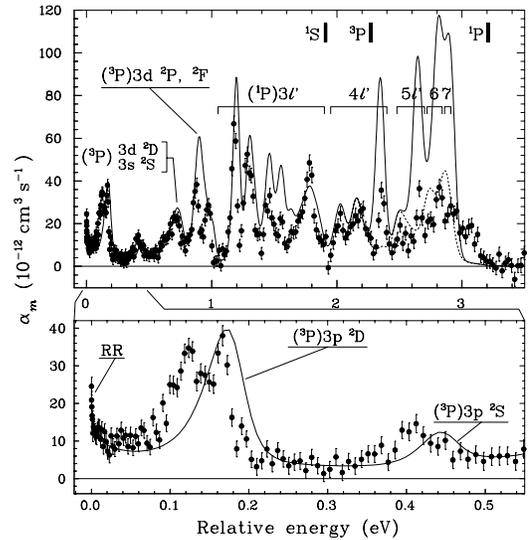}}
\vspace{1mm}
\caption[]{Recombination rate coefficient $\alpha_m(E)$ for metastable
  Li$^{+}\,(1s2s\,^3S)$ ions.  Dots: experimental data, obtained
  assuming an initial metastable fraction of $f^0_m=0.19$.  Full line:
  theory for $n_{c}=7$; labels refer to configurations
  $1s2p(^{1,3}P)n'l'$ and to excitation thresholds from the $2\,^3S$
  state.  Dashed line: theory omitting all states with $l'\geq3$.}
\label{ms}
\end{figure}

\samepage

\end{document}